\documentclass[aps,prd,notitlepage]{revtex4-1}

\usepackage{amsmath,amssymb}
\usepackage[bookmarks=false,colorlinks=true]{hyperref}
\usepackage{latexsym}
\usepackage{bm}

\newcommand{\Hycal}{\tilde{\mathcal{H}}_0}

\newcommand{\Hcoul}{\mathcal{H}_\gamma}
\newcommand{\Hgencoul}{{\mathcal{H}}^\text{gen}_{\gamma}}

\newcommand{\Hstark}{\mathcal{H}_{\gamma,F}}
\newcommand{\Hgenstark}{\mathcal{H}^\text{gen}_{\gamma,F}}

\newcommand{\Hgamma}{\mathcal{H}_{\gamma_1,\gamma_2}}
\newcommand{\Hgengamma}{\mathcal{H}^\text{gen}_{\gamma_1,\gamma_2}}

\numberwithin{equation}{section}

\newcommand{\be}{\begin{equation}}
\newcommand{\ee}{\end{equation}}

\newcommand{\bea}{\begin{eqnarray}}
\newcommand{\eea}{\end{eqnarray}}

\newcommand{\tq}{\tilde{q}}

 \newcommand{\pdu}{\mathcal{\pi}}

\begin{document}
\title{Integrability  and separation of variables in Calogero-Coulomb-Stark and two-center Calogero-Coulomb  systems}
\author{Tigran Hakobyan}
\email{tigran.hakobyan@ysu.am}

\author{Armen Nersessian}
\email{arnerses@ysu.am}
\affiliation{Yerevan State University, 1 Alex Manoogian Street, Yerevan, 0025, Armenia}
\affiliation{Tomsk Polytechnic University, Lenin Avenue 30, 634050 Tomsk, Russia}

\begin{abstract}
We propose the integrable  $N$-dimensional Calogero-Coulomb-Stark and two-center
Calogero-Coulomb systems and
construct  their  constants of motion via the Dunkl operators.
Their Schr\"odinger equations decouple in parabolic and elliptic coordinates
into the set of three differential equations like for the Coulomb-Stark and two-center Coulomb
problems. The Calogero term preserves the energy levels, but changes their
degrees of degeneracy.
\end{abstract}

\maketitle

\section{Introduction}

One of the most important features of the Coulomb problem is its maximal superintegrability  caused by the  conservation of the Runge-Lenz
vector.
As a consequence, its Hamiltonian admits separation of variables  in several coordinate systems,
 and  hence, a few types of integrable perturbations  with separation of variables.
  The textbook examples of such perturbations are (see, for example, Ref.~\onlinecite{ll}):
 \begin{enumerate}
 \item
 The Coulomb problem in constant electric field (the Coulomb-Stark problem) with  the Stark  potential
 $\delta V(\bm{r})=\bm{F}\cdot\bm{r}$,  which admits a separation of variables in parabolic coordinates.
 \item
 The two-center Coulomb problem, which admits a separation of variables in elliptic coordinates.
 \end{enumerate}
 These systems remain integrable despite the partially broken rotational symmetry along the highlighted
 direction.
 The latter  coincides  with the direction of the electric field $\bm{F}$ for the Coulomb-Stark problem, and  with  the vector $\bm{a}$  connecting
 two Coulomb charges for  the two-center Coulomb problems.
 As a result, the orthogonal angular momentum components and
the modified longitudinal component of the  Runge-Lenz vector are preserved.

However, the aforementioned systems are not solvable exactly.
In the  Coulomb-Stark problem one can get analytically only the perturbative spectrum, while
in the two-center Coulomb  system, the energy spectrum can be constructed only numerically,
except for some special cases \cite{komarov}. Nevertheless, the separation of variables  is crucial in the
 study of these systems.

Recently, together with O.~Lechtenfeld, we have observed that the $N$-dimensional Coulomb problem deformed
 by the rational Calogero potential  \cite{calogero0} (and  by its generalization associated with arbitrary root system \cite{perelomov})
 remains superintegrable and has the same energy spectrum as the original Coulomb model
\cite{CalCoul}.
The  Calogero-Coulomb Hamiltonian  is given by the expression
\be
 \label{int-coul}
\Hcoul=\sum_{i=1}^N \frac{{p}_i^2}{2}
+\sum_{i< j}^N
\frac{g(g-1)}{(x_i-x_j)^2} -\frac{\gamma}{\sqrt{\sum_ix_i^2}},
\ee
In Ref. \cite{Runge} we  have revealed an explicit expression for the analog of the Runge-Lenz vector in this system. We also observed  that being
formulated in terms of the Dunkl operators \cite{dunkl}, the  conserved quantities and  their algebra
 look pretty  similar to those in the conventional Coulomb problem.
 This led  us to the claim that most of the properties of the conventional Coulomb problem and
 its integrable perturbations  have their Calogero-Coulomb counterparts. 

 Thus, one may ask the following:

\smallskip

 \emph{Are  the high-dimensional Coulomb-Stark  and two-center  Coulomb problems with the Calogero potential
 (we will refer them as Calogero-Coulomb-Stark and two-center Calogero-Coulomb problems) integrable systems?
If they are, do they admit separation of variables, at least partial?}
\smallskip

 In this paper we  give positive answer  to both questions:

 i) The proper choice of the direction of the vectors $\bm{F}$ and $\bm{a}$
 makes the related systems  integrable. The corresponding  Hamiltonians are given by the expressions
 \be
 \label{int-stark}
 \Hstark=\sum_{i=1}^N \frac{{p}_i^2}{2}
+\sum_{i<j}^N\frac{g(g-1)}{(x_i-x_j)^2} -\frac{\gamma}{\sqrt{\sum_{i}x_i^2}}+\frac{F}{\sqrt N} {\sum_{i=1}^N x_i}
 \ee
 and
 \be
 \label{int-two}
 \Hgamma=\sum_{i=1}^N \frac{{p}_i^2}{2}
+\sum_{i<j}^N\frac{g(g-1)}{(x_i-x_j)^2} - \frac{\gamma_1}{\sqrt{\sum_i (x_i- a/\sqrt{N})^2}}-\frac{\gamma_2}{\sqrt{\sum_i
(x_i+a/\sqrt{N})^2}}.
  \ee
The constants of motion are the Dunkl-operator deformations of the corresponding integrals  in the underlying  Coulomb systems with the
subsequent symmetrization over the coordinates.

Note that  the integrability of the Coulomb-Stark and two-center Coulomb systems
(even without the Calogero interaction, even at the classical level)
does not follow from the existence of the Runge-Lenz vector or
the properties of the symmetry algebra of the usual Coulomb problem.
The preservation of the constructed constants of motion
in the systems \eqref{int-stark} and \eqref{int-two}  will be   verified
independently, based on the properties of the  Dunkl operators.

ii)
After the transition to the Jacobi (center-of-mass) coordinates the above listed systems admit
a complete separation of variables in parabolic/elliptic coordinates for  $N=2,3$
and a partial separation  for  $N>3$.
In fact, the Schr\"odinger equation decouples into
three parts, only  one of which depends on the  Calogero inverse-square term.
 The latter can be treated  as a deformation of the Schr\"odinger equation for the $SO(N-1)$ angular momentum,
 usually referred to as an angular Calogero Hamiltonian \cite{hny,sphCal,flp,fh}.

%

\smallskip

 The paper is organized as follows:

 In { Section 2} we describe the Calogero-Coulomb problem and its symmetry algebra, which
 contains the deformations of the angular momentum and Runge-Lenz vector by means of
 the Dunkl operator \cite{CalCoul,Runge}.
 Then we  perform the orthogonal transformation to the Jacobi coordinates and show that the transition
 to the spherical coordinates factorizes original Schr\"odinger equation into three  parts.
The first two  are just the radial and angular Schr\"odinger equations of the standard Coulomb problem.
 The third part produces the Schr\"odinger equation of the angular  Calogero model with the
 excluded center of mass, which has been solved in Ref.~\onlinecite{flp}.

 In { Section 3} we introduce the Calogero-Coulomb-Stark problem and  present explicit expressions
 for its constants of motion. They include the components of the Dunkl angular momentum,
 which are orthogonal to the direction of the electric field $\bm{F}$
 and the modified component  of the Runge-Lenz vector  along the field direction.
 We go on to show that after the transition to the Jacobi coordinates, the system possesses
a partial separation of variables in parabolic coordinates  as in  spherical coordinates.
Namely,  the Schr\"odinger equation  decouples into three  parts, where
the first two coincide with the respective  Schr\"odinger equations of the standard Coulomb
problem in parabolic coordinates, while the third  one
produces the Schr\"odinger equation of the angular  Calogero model with the
 excluded center of mass \cite{flp}.

 In { Section 4}  we  introduce the two-center Calogero-Coulomb problem and perform the  similar study for it.
 Again, the  components of the Dunkl angular momentum,
 orthogonal to the symmetry axis of the system given now by the vector  $\bm{a}$, are preserved.
 The last integral is a deformation of similar construction applied for the underlying two-center Coulomb system
 \cite{2centerRL}. The latter is based on the peculiar mixture of the longitudinal components of the
 two Runge-Lenz vectors, corresponding to each charge.
 After transition to the Jacobi coordinates, the system
 acquires a partial separation of variables in elliptic coordinates.

\section{Calogero-Coulomb problem}

\subsection{Coulomb problem}
The $N$-dimensional Coulomb problem is defined by the Hamiltonian
\be
\label{coul0}
\mathcal{H}_\gamma^0 = \frac{\bm{p}^2}{2}
-\frac{\gamma}{r}
\qquad
\text{with}
\qquad
r=\sqrt{\bm{x}^2}.
\ee
It possesses the maximum number ($2N-1$)  of functionally independent constants  of motion.
First, due to the  rotational invariance,  the system possesses conserving
 angular momentum tensor
\be
\label{L}
L_{ij}=x_i p_j-x_j p_i.
\ee
Its components satisfy the standard commutation rules for the $SO(N)$ generators:
\be
\label{comLL}
[L_{ij},L_{kl}]=\imath\delta_{lj}  L_{ik}+ \imath\delta_{ki} L_{jl} -\imath\delta_{kj} L_{il} - \imath \delta_{li} L_{jk}.
\ee
The angular momentum provides the system with
$N-1$ commuting quadratic (in momentum) constants of motion
\be
\bm{L}^2_k=\sum_{i<j\le k} L_{ij}^2
\qquad
\text{with}
\qquad
2\le k\le N,
\qquad
\bm{L}^2_N=\bm{L}^2.
\ee
They correspond to  the quadratic Casimir elements, constricted from the generators of
the sequence of the natural embeddings
$$
SO(2)\subset\dots\subset SO(k) \subset \dots \subset SO(N).
$$
Together with the Hamiltonian \eqref{coul0}, these constants of motion form the complete
set of the Liouville integrals.

Apart from the angular momentum, the $N$-dimensional Coulomb  motion
preserves the Runge-Lenz vector defined by expression
\be
\label{Bi}
A_i=\frac12 \sum_j\left\{L_{ij},p_j\right\}-\frac{\gamma x_i}{r},
\ee
where the curly braces mean the anticommutator:
$\{a,b\}=ab+ba$.

The angular momentum tensor and the Runge-Lenz vector define  the complete set of constants of motion
for the Coulomb system.

\subsection{Calogero-Coulomb Hamiltonian}
The Calogero-Coulomb problem introduced  in Ref.~\onlinecite{khare}  contains the additional
inverse-square  interaction term {\it a la} Calogero, as already mentioned in the Introduction.
It is a mixture of the $N$-particle rational Calogero model [see Refs.~\onlinecite{poly-rev,perelomov} for review] and  of the $N$-dimensional
Coulomb system.
In vector notations the Hamiltonian \eqref{int-coul} can be rewritten  as follows
\be
\Hcoul= \frac{\bm{p}^2}{2}
+\sum_{i< j}^N
\frac{g(g-1)}{(x_i-x_j)^2} -\frac{\gamma}{r}.
\label{coul}
\ee
It inherits most of the properties of the original Coulomb system \eqref{coul0}
and possesses hidden symmetries given by an analog of Runge-Lenz vector \cite{Runge}.
It is convenient to describe this system by means of the Dunkl operators
which makes transparent its analogy with the Coulomb problem.
Namely, instead of the $\Hcoul$, let us consider the more general  Hamiltonian
\be
\label{gencoul}
\Hgencoul=
\frac{\bm{\pdu}^2}{2}- \frac{\gamma}{r}=
\frac{\bm{p}^2}{2}
+\sum_{i< j} \frac{g(g-s_{ij})}{(x_i-x_j)^2} -\frac{\gamma}{r},
\ee
where the momentum modified by
 the Dunkl exchange operators \cite{dunkl} is used,
 \begin{subequations}
\be
\label{nabla}
\bm{\pdu} = -\imath\bm{\nabla},
\qquad
\nabla_i=\partial_i-\sum_{j\neq i}\frac{g}{x_i-x_j}s_{ij}.
\ee
The operator $s_{ij}$ in Eqs.~\eqref{gencoul} and \eqref{nabla} interchanges
the $i$-th coordinate with the $j$-th coordinate,
\be
\label{sij}
s_{ij}\Psi(\ldots, x_i,\ldots, x_j, \ldots) =\Psi(\ldots, x_j,\ldots, x_i, \ldots).
\ee
\end{subequations}
On the symmetric wavefunctions the generalized Hamiltonian $\Hgencoul$ reduces to the
Calogero-Coulomb Hamiltonian \eqref{coul}.

The Dunkl operators commute mutually like ordinary partial derivatives.
However, their commutations with coordinates
are nontrivial deformations of the Heisenberg algebra relations \cite{dunkl},
\begin{subequations}
\be
\label{com}
[\pdu_i,\pdu_j]=0,
\qquad
[\pdu_i, x_j]=-\imath S_{ij}.
\ee
Here the operators $S_{ij}$ are related with the permutations in the following
way:
\be
\label{Sij}
 S_{ij}=
 \begin{cases}
-g s_{ij} & \text{for}\; i\ne j,
\\
1+g\sum_{k\ne i} s_{ik}&\text{for}\; i=j.
 \end{cases}
\ee
\end{subequations}
Using the above relations,  it can be verified, that the operator-valued tensors $[S_{ij},x_k]$ and
$[S_{ij},\pdu_k]$ are  symmetric in all three indexes \cite{fh},
\be
\label{comSx}
[S_{ik},x_j]=[S_{jk},x_i],
\qquad
[S_{ik},\pdu_j]=[S_{jk},\pdu_i].
\ee
It must also be mentioned  that  the symmetries of the two-dimensional Calogero-Coulomb problem
associated with the dihedral group $D_2$ have been
recently studied in Ref.\cite{vinet} in terms of the Dunkl operators.

\subsection{Constants of motion for $\Hgencoul$}
Let us define the deformed angular momentum operator via the Dunkl momentum  \cite{kuznetsov,feigin,fh}:
\be
L_{ij}=x_i\pdu_j-x_j\pdu_i.
\label{Mij}
\ee
Its components  satisfy the following commutation relations \cite{fh},
\be
\label{comMM}
[L_{ij},L_{kl}]=\imath L_{ik}S_{lj} + \imath L_{jl}S_{ki}
 - \imath L_{il}S_{kj} - \imath L_{jk}S_{li}.
\ee
The Dunkl angular momentum inherits many properties of the ordinary angular momentum \eqref{L}.
In particular, the equations
\be
\sum_k\{S_{ik},\pdu_k\}=2\pdu_i,
\qquad
\sum_k\{S_{ik},x_k\}=2x_i,
\ee
which follow from the relations \eqref{com} and \eqref{Sij},
imply  that the operators $L_{ij}$ commute  with the Dunkl momentum
square and with the radius vector length \cite{fh}:
\be
\label{com-duLij}
[\bm{\pdu}^2,L_{ij}]=[r,L_{ij}]=0.
\ee
Hence, the Dunkl angular momentum is a constant  of motion of
the generalized Calogero-Coulomb Hamiltonian,
\cite{CalCoul}:
 \be
 \label{constMij}
[\Hgencoul,L_{ij}]=0.
\ee
The deformation of the Runge-Lenz vector \eqref{Bi}  is given by the expression  \cite{Runge}
\be
\label{Ai}
 A_i=\frac12 \sum_j\left\{ L_{ij},\pdu_j\right\}
+\frac\imath2[\pdu_i,S] -\frac{\gamma x_i}{r}.
\ee
It contains the permutation-group invariant element which vanishes in the absence of the Calogero term ($g=0$)
\be
\label{S}
S=\sum_{i<j}S_{ij}:
\qquad
[S,S_{ij}]=0.
\ee

The deformed Runge-Lenz vector is also a constant of motion:
\be
\label{constAi}
[\Hgencoul, A_i]=0.
\ee
The proof is given in Ref.~\onlinecite{Runge}, and we repeat it here for the completeness.

According to \eqref{com-duLij}, the commutator  of
the Hamiltonian with the first term from the rhs of Eq.~\eqref{Ai}
is proportional to
\be
\Big[\sum_{j}\{L_{ij},\pdu_j\}\,,\,\frac{\gamma}{r} \,\Big]
=  -\frac{\imath\gamma}{r^3}  \sum_j \left\{L_{ij},x_j\right\}
 =   \Big\{\frac{\imath\gamma}{r}, \pdu_i\Big\}- \imath\gamma\sum_j \left\{ \frac{ x_ix_j}{r^3}, \pdu_j\right\},
\label{AHpr1}
\ee
and its commutator with the last term is proportional to
  \begin{gather}
 \label{AHpr2}
 \begin{split}
\Big[\bm{\pdu}^2, \,\frac{x_i}{r} \Big]
 =    \imath\sum_j\Big\{ \frac{x_ix_j}{r^3}-\frac{S_{ij}}{r},\pdu_j\Big\}.
 \end{split}
  \end{gather}
Combining the relations \eqref{AHpr1} and \eqref{AHpr2}
and using the definition \eqref{S},
we obtain:
\be
\label{final}
\Big[ \frac12\sum_{j}  \{L_{ij},   \pdu_j\}-\frac{\gamma x_i}{r}\,, \, \Hgencoul\Big]
=  \imath \sum_{j}\left\{\frac{\gamma S_{ij}}{2r}, \pdu_i-\pdu_j\right\}
=\frac\imath2\big[ [S,\pdu_i],\Hgencoul\big].
\ee
In the last equation we used the identity
\[
\sum_j(\pdu_j-\pdu_i)S_{ij}=[S,\pdu_i].
\]
Finally, Eq.~\eqref{final} completes the proof of the conservation condition \eqref{constAi}.\\

The deformed Runge-Lenz vector  can be written  directly through the coordinate and Dunkl momentum
by avoiding
the invariant $S$:
\be
\label{Avec}
\bm{A}=\bm{x}\left(\bm{\pdu}^2-\frac{\gamma}{r}\right) -\left(rp_r+\frac{N-1}{2\imath}\right)\bm{\pdu}.
\ee
Here $p_r$ is the ordinary radial momentum canonically conjugate to the radial coordinate:
$[p_r,r]=-\imath $.
It turns into its $g=0$ analog under the replacement $\bm{\pdu}\to\bm{p}$.

The relation \eqref{Avec} can be checked by inserting  Eq.~\eqref{Mij} into Eq.~\eqref{Ai} and
applying the identity
\be
\label{pdu-x}
\bm{\pdu}\cdot \bm{x}+\bm{x}\cdot \bm{\pdu}= 2(\bm{x}\cdot \bm{p})-\imath N=2r p_r -\imath N.
\ee
This identity  is a direct consequence of the deformed canonical commutation relations
\eqref{com}.

The commutation relations among the Dunkl angular momentum components
\eqref{comMM} are supplemented by those  which incorporate
the deformed Runge-Lenz  momentum and Hamiltonian \cite{Runge}:
\begin{gather}
\label{comA}
[A_i,L_{kl}]=\imath A_kS_{li} - \imath A_lS_{ki},\qquad [A_i,A_j]=-2\imath \Hgencoul L_{ij}.
\end{gather}
In the absence of the inverse-square interaction, the exchange operator $S_{ij}$
is reduced to the Kronecker delta, and we get the familiar algebra, formed by the
symmetry generators of the Coulomb problem.
On a negative constant-energy surface $\Hcoul^0=E<0$ it reduces to
the orthogonal algebra $so(N+1)$, and on the positive one, $\Hcoul^0=E>0$, to the $so(N,1)$.

\subsection{Constants of motion for $\Hcoul$}
Let us remember  that the Calogeo-Colomb problem can be  obtained by the restriction of the generalized
Hamiltonian \eqref{gencoul} to symmetric wavefunctions \eqref{sij}.
Therefore, its constants of motion can be constructed by taking the symmetric polynomials
depending on the components of the Dunkl angular momentum and Runge-Lenz vector \cite{CalCoul,Runge}:
\begin{gather}
\label{M2k}
\mathcal{L}_{2k}=\sum_{i<j}L_{ij}^{2k},
\\
\label{Ak}
 {\cal A}_k=\sum_i A_i^k.
\end{gather}
The expressions above demonstrate that the  Calogero-Coulomb problem is a superintegrable system, like the pure Calogero \cite{woj83} and
Coulomb models.
Among these  integrals  only the two    are  quadratic on momenta.
The first one is the square of the Dunkl angular momentum,
\be
\label{L2-I}
\mathcal{L}_2= 2\mathcal{I} + S(S-N+2),
\ee
where $ \mathcal{I}$ is the angular part of the Calogero (Calogero-Coulomb) Hamiltonian \cite{fh}:
\be
\label{Hsph}
\Hcoul=\frac{p_r^2}{2} -\imath\frac{N-1}{2r}p_r-\frac{\gamma}{r}+\frac{\mathcal{I}}{r^2}.
\ee
It coincides with the Casimir element of the Dunkl angular momentum algebra and
is also a conserved quantity:
\be
\label{casimir}
[ L_{ij},\mathcal{I}]=0,
\qquad
[\mathcal{I},\Hcoul]=0.
\ee
So it can be used instead of $\mathcal{L}_2$ as a constant of motion.

The second quadratic integral is given by the sum of the components of the
deformed Runge-Lenz vector
defined in Ref.~\cite{CalCoul}. For  convenience, we take a rescaled expression,
\be
\label{A1}
\frac{1}{\sqrt{N}}{\cal A}_1 =
\frac{1}{\sqrt{N}} \sum_j\left\{ L_{ij},\pdu_j\right\}
-\frac{\gamma x_0}{r}
=
x_0\Big(2\Hgencoul +\frac{\gamma}r\Big)
- \Big(rp_r + \frac{N-1}{2\imath}\Big)p_0.
\ee
It depends on the normalized center-of-mass coordinate
and momentum
\be
\label{x0}
x_0=\frac{1}{\sqrt{N}}\sum_{i=1}^N x_i,
\qquad
p_0=\frac{1}{\sqrt{N}}\sum_{i=1}^N p_i.
\ee
The existence of two second-order integrals suggests that the  Calogero-Coulomb problem
 admits a separation of variables for  dimension  $N\le3$.

\subsection{Jacobi coordinates}
The  Jacobi coordinates  \cite{reed78,hny,hln12} guarantee the separations of
the  center-of-mass \eqref{x0}  from the relative motion, keeping
the kinetic term's shape.
They are defined  in the following way:
\begin{subequations}
\be
\label{jacobi}
y_0=x_0=\frac{1}{\sqrt{N}}(x_1+\dots +x_N),
\qquad
 y_k=\frac{1}{\sqrt{k(k+1)}}(x_1+\dots +x_k-kx_{k+1}), \quad 1\le k\le N-1.
\ee
The first coordinate describes the center of mass, while the
other coordinates characterize the relative motion. It can be verified that the above
transformation is  \emph{orthogonal}:
\be
\label{Okl}
y_k=\sum_{l=1}^N O_{kl}x_l,
\qquad
x_k=\sum_{l=0}^{N-1}O_{lk}y_l.
\ee
\end{subequations}
In order to express the Hamiltonian and Dunkl momentum in terms of the Jacobi coordinates,
we note that the permutation operators are  just  orthogonal reflections  across
the hyperplanes $\bm{\alpha}\cdot \bm{x}=0$
 in the $N$-dimensional coordinate space:
\be
\label{s-alpha}
s_{ij} \bm{x}=s_{\bm{\alpha}} \bm{x}
:= \bm{x}- \frac{2(\bm{\alpha}\cdot \bm{x})}
{\bm{\alpha}\cdot\bm{\alpha}}\bm{\alpha}
\qquad
\text{with}
\qquad
\bm{\alpha}=\bm{\alpha}_{ij}=\bm{e}_i-\bm{e}_j,
\ee
where $\bm{e}_i$ form the standard basis in $\mathbb{R}^N$:
$\bm{x}=\sum_i x_i\bm{e}_i$.
The operators $S_{ij}$ are expressed in a similar manner.
Note that this representation
admits the generalization of the exchange operators \eqref{nabla} to an arbitrary
finite reflection group known also as a Coxeter group \cite{dunkl}.  The vector $\bm{\alpha}$
is called a root of the Coxeter system. A particular choice of roots in Eq.~\eqref{s-alpha}
defines  the $A_{N-1}$ Coxeter system.
They separate the center-of mass coordinate from the relative motion.

We rewrite the reflections \eqref{s-alpha} in the Jacobi coordinates using the vector notations:
\begin{subequations}
\be
\label{s-beta}
s_{ij} \bm{y}=s_{\bm{\beta}} \bm{y}
= \bm{y}- \frac{2(\bm{\beta}\cdot \bm{y})}
{\bm{\beta}\cdot\bm{\beta}}\bm{\beta}.
\ee
Here the  vectors $\bm{\beta}$ are images of the corresponding root vectors $\bm{\alpha}$
under the Jacobi transformation \eqref{Okl}:
\be
\beta_k = \sum_{k'} O_{kk'}\alpha_{k'},
\qquad
\beta_{ij,k}={O_{ki}-O_{kj}}.
\ee
\end{subequations}
Their explicit form  \eqref{jacobi} ensures the relation
$\beta_{ij,0}=0$. In fact,  the reflection \eqref{s-beta} is restricted
to the relative coordinates (see Eq.~\eqref{rel} below).

In Jacobi coordinates the inverse-square Calogero potential in the Hamiltonian \eqref{coul} reads
\be
\label{coxeter}
V (\bm{y})= \sum_{i<j}
\frac{g (g-1)}{(\bm{\alpha}_{ij}\cdot \bm{x})^2}
=
\sum_{i<j}\frac{g (g-1)}{(\bm{\beta}_{ij}\cdot \bm{y})^2}.
\ee
Evidently, the Dunkl momenta along the new coordinate directions  are related to the
usual Dunkl momenta
$\pdu_{x_i}=\pdu_i$ by the vector transformation
\be
\tilde\pdu_i=\pdu_{y_i}=\sum_{i'=1}^N O_{ii'}\pdu_{i'}.
\ee
Its zeroth component describes the center-of-mass momentum \eqref{x0}:
$\tilde\pdu_0=\tilde p_{0}=p_0$.
In order to ensure the invariance of the defining relations  \eqref{com},
the operators $S_{ij}$ must behave as a second-order tensor in an orthogonal
transformation:
\be
\label{com-tilde}
[\tilde\pdu_i, \tilde\pdu_j]=0,
\qquad
[\tilde\pdu_i,  y_j]=-\imath \tilde S_{ij}=-\imath \sum_{i',j'=1}^N O_{ii'}O_{jj'}S_{i'j'}.
\ee

In Jacobi coordinates, the Cherednik algebra decouples into two commuting parts describing
 the center-of-mass ($y_0=x_0$, $p_0$) and  relative ($y_i$, $\tilde\pi_i$ with $1\le i\le N-1$) motions, respectively.
Indeed, exploiting the defining relations \eqref{Sij} and the orthogonality of the Jacobi transformation \eqref{Okl}, we obtain
\be
\label{Si0}
\tilde S_{i0}=\frac{1}{\sqrt{N}}\sum_{j'=1}^N S_{i'j'}\sum_{i'=1}^N O_{ii'}=\sum_{i'=1}^N O_{ii'}O_{0i'}=\delta_{i0}.
\ee
It implies
\be
\label{diy0}
[\tilde\pdu_i, y_0]=[\tilde\pdu_i, x_0]=[p_0, y_i]=-\imath \delta_{i0}.
\ee
Note that the constraint \eqref{Si0} is equivalent in old coordinates to the condition $\sum_j S_{ij}=1$,
which follows from the defining relation \eqref{Sij}.

Below  we use  the following vector notations for the description
of the relative coordinates and momenta:
\be
\label{rel}
\bm{y}= (y_1,\dots, y_{N-1}),
\qquad
\tilde{\bm{\pdu}} = (\tilde\pdu_1,\dots \tilde\pdu_{N-1}).
\ee
Note that the operators $\tilde S_{ij}$ do not permute the coordinates any more.
However,  the tensors $[\tilde S_{ij},y_k]$ and
$[\tilde S_{ij},\tilde\pdu_k]$ still remain symmetric so that the relations \eqref{comSx}
are preserved:
\be
\label{Spi}
[\tilde S_{ik},y_j]=[\tilde S_{jk},y_i],
\qquad
[\tilde S_{ik},\tilde\pdu_j]=[\tilde S_{jk},\tilde\pdu_i].
\ee
This property persists for   any Coxeter system \cite{fh}.

Of course, similarly to the operators $S_{ij}$, the Dunkl angular momentum transforms as
a second-order tensor under an orthogonal transformation:
\begin{subequations}
\be
\label{Lij-tilde}
\tilde L_{ij} = L_{y_iy_j}=\sum_{i',j'=1}^N O_{ii'}O_{jj'}L_{i'j'}.
\ee
The rotated components obey the commutation relations
\eqref{comMM} with the permutation generators $S_{ij}$ replaced by
their shifted counterparts $\tilde S_{ij}$.
 They acquire the standard form  \eqref{Mij} in the Jacobi coordinates and momenta,
\begin{align}
\label{Lij-jac}
\tilde L_{ij}&= y_i\tilde{\pdu}_j -  y_j\tilde{\pdu}_j
 \qquad
 \text{with}
 \qquad
 0\le i,j\le N-1,
\\
\label{L0j}
 \tilde L_{0i}&= x_0\tilde{\pdu}_i - y_jp_0.
\end{align}
\end{subequations}
Using \eqref{Si0}, we can rewrite them
in the form which splits up the center-of-mass and relative degrees of freedom:
\begin{subequations}
\begin{align}
\label{comMM-tilde}
[\tilde L_{ij},\tilde L_{kl}]&=\imath\tilde  L_{ik}\tilde S_{lj} + \imath\tilde L_{jl}\tilde S_{ki}
 - \imath\tilde L_{il}\tilde S_{kj} - \imath \tilde L_{jk}\tilde S_{li},
\\
\label{comMM0i}
[\tilde L_{ij},\tilde L_{0l}]& = \imath \tilde L_{0j}\tilde S_{li}-\imath\tilde  L_{0i}\tilde S_{lj} ,
\\
\label{comMM00}
[\tilde L_{0j},\tilde L_{0l}]&= \imath\tilde L_{jl}
\qquad
\text{with}
\qquad
1\le i,j,k,l\le N-1.
\end{align}
\end{subequations}
So the  generators \eqref{Lij-jac} form the deformed analog of the $so(N-1)$
angular momentum algebra consisting of the relative degrees of freedom.

Usually, in the Coulomb problem, the separation of variables is performed in spherical coordinates.
However, for the (partial) separation of variables in spherical coordinates, we should first perform
the orthogonal transformation to the Jacobi coordinates
in order to extract the center of mass \cite{hny}.

\subsection{Separation of variables in  spherical coordinates}
Let us extract now the coordinates of  the center of mass from the Calogero-Coulomb  Hamiltonian \eqref{coul},
\be
\label{decomp}
\Hcoul= \frac{p_ 0^2}{2}
-\frac{\gamma}{\sqrt{x^2_0+y^2}} +\Hycal
\qquad
\text{with}
\qquad
y=\sqrt{\bm{y}^2}.
\ee
The last term is just the  Calogero Hamiltonian with the excluded center of mass,
which depends on the $N-1$ relative degrees of freedom. In the Jacobi coordinates, it reads:
\be
\label{n-1}
\Hycal=\frac{\tilde{\bm{p}}^2}{2}+
\sum_{i<j}\frac{g(g-1)}{(\bm{\beta}_{ij}\cdot \bm{y})^2} .
\ee
For the simplest $N=2$ case, the relative Calogero model is reduced to a one-dimensional
conformal mechanical system with the Hamiltonian
\begin{subequations}
\be
\Hycal=\frac{\tilde p^2}{2}+\frac{g(g-1)}{2y^2}.
\label{A-1}
\ee
In the  $N=3$ case, it takes more complicated  form:
\be
\Hycal=
\frac{\tilde p_1^2+\tilde p_2^2}{2}  +
 \frac{g(g-1)}{2y^2_1}+\frac{2g(g-1)}{(\sqrt{3}y_2+y_1)^2}+\frac{2g(g-1)}{(\sqrt{3}y_2-y_1)^2}.
\label{A-2}
\ee
\end{subequations}

In the general case, due to the conformal symmetry,
 the relative Hamiltonian in \eqref{decomp} can be spit further  into its radial and angular parts,
 \be
\Hycal=\frac{\tilde{\bm{p}}^2}{2}-\frac{\tilde{\cal I}(\varphi_{\iota},p_{\varphi_\alpha})}{y^2},
 \label{angN-1}
 \ee
 where
 $\varphi_\iota $ are the angular coordinates, parametrizing the
 $(N-2)$-dimensional sphere,
 and $\tilde{\cal I}$ is the Hamiltonian of the  angular Calogero model with the excluded center of mass
 (relative angular Calogero model).
 The latter system has been investigated in detail in the series of papers \cite{hny,sphCal,flp}.
For   the $N=3$ case  it is reduced to a
one-dimensional exactly solvable system on circle, introduced by Jacobi in the middle of the 19th century
\cite{jacobi}:
\be
\tilde{\cal I}=
\frac{p_\varphi^2}{2}+\frac{9g(g-1)}{2\cos^2 3\varphi}.
\label{2ph}
\ee
For higher dimensions,  the relative angular Calogero Hamiltonian has more complicate form
(see Refs.~\onlinecite{hny,cor-lech}).

 Now we are ready to transition  to  spherical coordinates.
 The azimuthal angle $\theta$ is chosen to measure the projection
 of the position vector to the relative coordinate's hyperplane:
 \be
 \label{sph}
x_0=r\cos\theta,
\qquad
\bm{y}=r\sin\theta\, \tilde{\bm{n}}(\varphi_\iota).
 \ee
 Here the $(N-1)$-dimensional unit vector $\tilde{\bm{n}}$
is parametrized by the remaining  angles $\varphi_\iota$ with $1\le\iota\le N-2$.
In these coordinates the Hamiltonian \eqref{Hsph} looks as follows:
\be
\label{coul-sph}
\Hcoul=- \frac{1}{2r^{N-1}}\frac{\partial}{\partial r}r^{N-1}\frac{\partial}{\partial r}
+  \frac{\mathcal{I}(\theta,\varphi_\iota,\partial_\theta,\partial_{\varphi_\iota})}{r^2}
    -\frac{\gamma}{r}.
\ee
The angular  Calogero Hamiltonian $\mathcal{I}$ is related with the
relative angular  Hamiltonian $\tilde{\cal I}$ by the expression
\be
\mathcal{I}(\theta,\varphi_\iota,\partial_\theta,\partial_{\varphi_\iota})
=\frac{\tilde{\cal I}(\varphi_\iota,\partial_{\varphi_\iota})}{\sin^2\theta}
   -\frac{1}{2\sin^{N-2}\theta}\ \frac{\partial}{\partial\theta} \sin^{N-2}\theta\ \frac{\partial}{\partial\theta}.
\ee
Both quantities $\mathcal{I}$ and $\tilde{\cal I}$ are constants  of motion.  They separate
the radial $r$ and azimuthal $\theta$ coordinates, each being symmetric in the primary coordinates $x_i$,
from the relative angular coordinates $\varphi_\iota$.
Let us choose  the factorized wavefunction:
\be
\label{factor}
\Psi(r,\theta,\varphi_\iota)=R(r)\Phi(\theta)\psi(\varphi_\iota).
\ee
It decouples the Schr\"odinger equation $\Hcoul\Psi=E\Psi $
into the following system of differential equations:
\begin{subequations}
\begin{align}
 \label{R-Shr}
\left(\frac{1}{r^{N-1}}\frac{d}{dr}r^{N-1}\frac{d}{dr}-
 \frac{l(l+N-2)}{r^2}+\frac{2\gamma}{r}+2E_n
\right)R_{n_rl} (r)&=0,
 \\
  \label{Phi-Shr}
 \left( \frac{1}{\sin^{N-2}\theta}\ \frac{d}{d\theta} \sin^{N-2}\theta\ \frac{d}{d\theta} -\frac{\tq(\tq+N-3)}{\sin^2\theta}+l(l+N-2)
\right)\Phi_l(\theta)& = 0,
 \\
 \label{I-Shr}
 \left(2\tilde{\cal I}(\varphi_\iota,\partial_{\varphi_\iota})-\tq(\tq+N-3)\right)\psi_{\tq}(\varphi_\iota)
&=0.
\end{align}
\end{subequations}
The last equation describes the spectrum and eigenstates of $\tilde{\cal I}$,
which were recently investigated in detail  in Ref.~\onlinecite{flp}.
In particular, the spectrum is determined by the numbers
\be
\begin{split}
\tq=\frac{gN(N-1)}{2} + 3l_3+\ldots +Nl_N
\qquad
\text{with}
\qquad
l_i=0,1,2,\dots.
\end{split}
\label{Iq}
\ee
For integer values of the coupling $g$, the angular energy spectrum  is that of a free particle with angular
momentum $\tq$ on the $(N-2)$-dimensional sphere, but it has a significantly lower degeneracy due to
the restriction to the symmetric wavefunctions  \cite{flp,cor-lech}.

The first two equations above are similar to those for the $N$-dimensional Coulomb problem.
The solutions of both equations can be found, for instance,  in Ref.~\onlinecite{NStark}.
The second equation \eqref{Phi-Shr} determines the spectrum and  eigenfunction of the angular Hamiltonian $\mathcal{I}$.
For integer coupling $g$, its spectrum coincides with  that of a free particle with angular
momentum $l$ on the $(N-1)$-dimensional sphere  \cite{flp}.
The first  equation \eqref{R-Shr} is related to the radial Schr\"odinger equation for the
$N$-dimensional Coulomb problem.

For the positive integer values of the coupling constant, $g=1,2,\dots$, the solutions
of Eq.~\eqref{Phi-Shr}  exist, provided the quantum number  $l$ takes the
values
\be
l=\tq+l_1
\qquad
\text{with}
\qquad
l_1=0,1,2,\dots.
\ee
so that  $l\ge\frac12N(N-1)g $.

Similarly, the well-defined solution of the radial equation \eqref{R-Shr} exists
for the energy spectrum depending on the principal quantum number   $n$:
\be
\label{En}
E_n=-\frac{\gamma^2}{2\left(n +\frac{N-3}{2}\right)^2},
\qquad
n=n_r+l+1,
\qquad  n_r=0,1,2,\dots.
\ee
Hence, the Calogero-Coulomb problem remains degenerate with respect to $q$.
The energy spectrum of the system is the same as in the $N$-dimensional Coulomb problem,
but with a lower degeneracy due to the restriction to the symmetric wavefunctions on the initial
coordinates.

\section{Coulomb-Calogero-Stark problem}

\subsection{Coulomb-Stark problem}
It is well known that the $N$-dimensional Coulomb problem in constant uniform electric field
$\bm{F}$ remains integrable and admits a separation of variables in parabolic coordinates. Its Hamiltonian  is defined by the expression
\be
\label{stark0}
\Hstark^0 = \frac{\bm{p}^2}{2}
-\frac{\gamma}{r}+\bm{F}\cdot\bm{x}.
\ee
The perturbative spectrum of this system can be found in  Refs. \cite{ll,NStark}.

The Stark interaction breaks the $SO(N)$ rotation symmetry
of the Coulomb problem down to the $SO(N-1)$ symmetry.
The corresponding constants of motion are given  by the components of
the angular momentum \eqref{L}
which are orthogonal to the direction of the electric field,
\be
\label{Lbot}
L^{\bot}_{ij}= L_{ij}+n_in_kL_{jk}-n_jn_kL_{ik},
\ee
Here $\bm{n}$
is the unit vector along the electric field:
\be
\label{n}
\bm{F}=F\bm{n},
\qquad
\bm{n}^2=1.
\ee
In  coordinates, where the unit vector $\bm{n}$ is the $N$-th  basic vector,
the perpendicular components   \eqref{Lbot} acquire  the form of the $SO(N-1)$ algebra
standard basis. Among them,  one can choose the $N-1$ commuting constants  of motion.

Apart from the angular momentum, the Coulomb-Stark system
possesses a constant of motion, which is inherited from the $N$-dimensional Runge-Lenz vector
\eqref{Ai}.
It reads
(see Ref.~\onlinecite{redmond} for the $N = 3$ case)
\be
\label{B}
 A=  \bm{n}\cdot\bm{A}
 -\frac {F}{2} \left(r^2-(\bm{n}\cdot\bm{x})^2\right)
\ee
and complements the Liouville integrals to the full set.

\subsection{Integrable Hamiltonian of the Coulomb-Calogero-Stark system}
The inclusion of the inverse-square Calogero potential breaks down the $SO(N)$ rotational
symmetry of the initial Hamiltonian \eqref{coul0}  to the discrete
group containing the coordinate permutations \eqref{sij}.
Therefore, the different field directions $\bm{n}$ become  nonequivalent.
Among them, there is a preferable direction, which ensures the permutation invariance
of the Stark term $\bm{F}\cdot\bm{x}$:
\be
\label{dir}
\bm{n}=\frac{1}{\sqrt{N}}(1,\ldots, 1).
\ee
This is virtually the direction of the center-of-mass coordinate, so that in the
Jacobi  system \eqref{jacobi}, the Stark interaction takes
the following form:
\be
\label{FyN}
\bm{F}\cdot\bm{x}=Fx_0.
\ee
So the generalized Calogero-Coulomb Hamiltonian in an external electric field is set to
\be
\label{genstark}
\begin{split}
\Hgenstark=\Hgencoul+Fx_0&=
\frac{\bm{\pdu}^2}{2}- \frac{\gamma}{r}+Fx_0
\\
&=\frac{\bm{p}^2}{2}
+\sum_{i< j} \frac{g(g-s_{ij})}{(x_i-x_j)^2} -\frac{\gamma}{r}+ F x_0.
\end{split}
\ee
We see that the particular choice of the field direction \eqref{dir}
keeps the entire Hamiltonian invariant under the coordinate permutations:
\be
[\Hgenstark\,,\,s_{ij}]=0.
\ee
Therefore, it is well defined on the symmetric wavefunctions, where
it is reduced to the Hamiltonian, as mentioned in the Introduction
[see Eq.~\eqref{int-stark}]. We rewrite it here in a more
compact form:
\be
\label{stark}
\Hstark=\frac{\bm{p}^2}{2}
+\sum_{i<j}^N\frac{g(g-1)}{(x_i-x_j)^2} -\frac{\gamma}{r}+F x_0.
\ee

\subsection{Constants of motion for $\Hgencoul$}
It turns out that  the symmetry generators  \eqref{Lbot}
and \eqref{B} to the Stark Hamiltonian \eqref{stark0}
are extended straightforwardly to the generalized Calogero-Coulomb-Stark problem.

First, consider the invariants of the generalized Calogero Hamiltonian $\Hgencoul$.
Substituting the highlighted direction \eqref{dir} into the orthogonal angular momentum tensor
\eqref{Lbot}, which describe the remaining $SO(N-1)$ symmetry, we immediately get
\be
\label{Mbot}
L^{\bot}_{ij}= L_{ij}+\frac{1}{N}\sum_k (L_{jk}-L_{ik}).
\ee
Of course, this definition remains valid for  the Dunkl angular momentum \eqref{Mij} too.
Below we will consider just this case.
It is easy to verify the following commutation relations among the coordinates
and Dunkl angular momenta:
\be
\label{comLx}
[L_{ij},x_l]= \imath x_j S_{il}-\imath x_i S_{jl},
\qquad
[L_{ij},x_0]=-\imath (x_i- x_j).
\ee
They ensure that the orthogonal components of the
Dunkl angular momentum are constants of motion of the generalized
Calogero-Coulomb-Stark problem:
\be
\label{comMH}
\big[{L}^{\bot}_{ij},\Hgenstark\big]=0.
\ee
Alternatively, one can switch to the relative Jacobi coordinates and  use the
basis formed by the $\tilde L_{ij}$ \eqref{Lij-jac}   for the description of the
same algebra \eqref{Mbot}. Using the established identity $S_{0i}=\delta_{0i}$
it is easy to ensure an analog of the commutation relations
\eqref{comLx}:
\be
[\tilde L_{ij}, y_l]= \imath y_j \tilde S_{il}-\imath y_i \tilde S_{jl},
\qquad
[\tilde L_{ij},y_0]=0.
\ee
The above relation implies
\be
\label{comMH-1}
\big[\tilde{L}_{ij},\Hgenstark\big]=0.
\ee
The deformed analog of the modified Runge-Lenz  component along the field direction
 is a constant of motion, too,
\begin{align}
\label{Apar}
A&=\frac{1}{\sqrt{N}}{\cal A}_1- \frac{F}{2}\left (r^2-x_0^2\right)
\\
&=x_0\Big(2\Hgenstark +\frac{\gamma}r\Big)
- \Big(rp_r + \frac{N-1}{2\imath}\Big)p_0  - \frac{F}{2}\left (r^2 + 3x_0^2\right).
\end{align}
where $\mathcal{A}_1=\sum_i A_i$ according to the definition \eqref{Ak}, and we have used
the identity  \eqref{A1}.

The proof is more complicated, and we present here some details of the derivation.
Using the definitions \eqref{genstark}, \eqref{Apar}, and the already proven fact that the
  deformed Runge-Lenz vector is a constant  of motion of the unperturbed Hamiltonian  [see \eqref{constAi}],
we obtain:
\be
\label{comAH1}
\big[A, \Hgenstark \big]=
\frac{F}{\sqrt{N}}[\mathcal{A}_1,x_0]+\frac{F}{4}\big[\pdu^2,\,r^2-x_0^2\big].
\ee
First, we calculate the second commutator from the right side of the above equation.
Using a simple algebra and the relations \eqref{pdu-x}, we get
\begin{subequations}
\begin{gather}
\label{comHr}
[\bm{\pdu}^2, r^2]
=-2\imath (\bm{\pdu}\cdot \bm{x}+\bm{x}\cdot\bm{\pdu})=-4\imath rp_r-2N,
\\
\label{comHx0}
[\bm{\pdu}^2, x_0^2]
=[p_0^2, x_0^2]=-4\imath x_0p_0-2.
\end{gather}
\end{subequations}
Next, applying the identities \eqref{A1} and  \eqref{comHx0},
the first commutator in Eq.~\eqref{comAH1} can be simplified as follows:
\be
\label{comAx0}
\begin{split}
\frac{1}{\sqrt{N}}[\mathcal{A}_1,  x_0]
&=x_0[ \bm{\pdu}^2,x_0]+\imath x_0p_0+\imath rp_r+\frac{N-1}{2}
\\
&=\imath( rp_r-x_0p_0)+\frac{N-1}{2} .
\end{split}
\ee
Finally, substituting the commutation relations  \eqref{comHr}, \eqref{comHx0},
and \eqref{comAx0} into the equation \eqref{comAH1}, we get the desired
conservation condition:
\be
\label{comAH}
[A, \Hgenstark ]=0.
\ee
It is easy to check that the modified Runge-Lenz  longitudinal invariant commutes with
the deformed angular momentum restricted to the relative coordinates:
\be
[\tilde{L}_{ij},A]=0
\qquad
\text{with}
\qquad
1\le i,j\le N-1.
\ee
So we have proved  that in the presence of a constant uniform electric field,
the generalized Calogero-Coulomb model \eqref{genstark} still remains an integrable system.
The integrals of motion are deformations of those for the conventional Coulomb
model with the Stark interaction by the Dunkl operators.

\subsection{Constants of motion of $\Hcoul$}
The integrals  of the pure Calogero-Coulomb system \eqref{stark} obtained by the
restriction to the symmetric wavefunctions, must be symmetric too. Since the  longitude  component
of the Runge-Lenz vector   \eqref{Apar} obeys this condition, it remains as a correct integral for this system,
\be
[A,\Hstark]=0.
\ee

We should take symmetric expressions of the kinematical constants of motion too, as
in the absence of the electric field \eqref{M2k}. For this purpose, it is more suitable to use the
angular momentum  in Jacobi coordinates \eqref{Lij-tilde}. Thus, we have
\be
[\Hstark,\tilde{\mathcal{L}}_{2k}]=0,
\qquad
\tilde{\mathcal{L}}_{2k}=\sum_{1\le i<j\le N- 1}\tilde L_{ij}^{2k}.
\ee
The first member of this set is the square of the relative Dunkl angular
momentum:
\be
\tilde{\mathcal{L}}_2= 2\tilde{\cal I} + S(S-N+2).
\ee
Here $ \tilde {\cal I}$ is the angular part of the Calogero Hamiltonian
with reduced center of mass (the relative angular Calogero Hamiltonian).
{
It also forms a single Casimir element of the
relative Dunkl angular momentum algebra \cite{fh}:
\be
\label{Hsph-rel}
[ \tilde{L}_{ij},\tilde{\cal I}]=0,
\qquad
\tilde{H}_0=\frac{p_y^2}{2} -\imath\frac{N-2}{2}p_y+\frac{\tilde{\cal I}}{y^2}.
\ee
}
Thus, we have proved  the integrability of the Calogero-Coulomb-Stark system.

\subsection{Separation of variables in parabolic coordinates}
It is well known that the Coulomb-Stark system \eqref{stark0} admits separation of variables in parabolic coordinates,
in which the Stark effect can be immediately calculated.
Is this true for the Calogero-Coulomb-Stark system?
Below we will show that the system admits complete separation of variables  in parabolic coordinates
for $N=2,3$ and  partial separation for $N>3$.

In the  Jacobi coordinates \eqref{jacobi}, the Calogero-Coulomb-Stark system acquires the following  form,
\be
\Hstark=\frac{p_0^2}{2}
-\frac{\gamma}{\sqrt{x^2_0+y^2}} +Fx_0 +\Hycal,
\label{stark-par}
\ee
where again the last term is  the  Calogero Hamiltonian with the reduced center of mass \eqref{n-1}.
From the spherical coordinates $(r,\theta,\varphi_\iota)$ \eqref{sph}, we move to the
parabolic coordinates  $(\xi,\eta,\varphi_\iota)$, using the map which leaves the relative angular variables $\varphi_\iota$
unaltered: 
\be
\label{par}
\xi=r+x_0,
\qquad
\eta=r-x_0
\qquad
\text{with}
\qquad
x_0=r\cos\theta.
\ee
The inverse transformation reads
\be
 \label{par-inv}
x_0=\frac{\xi-\eta}{2},
\qquad
r=\frac{\xi+\eta}{2},
\qquad
\bm{y}=\sqrt{\xi\eta} \,\tilde{\bm{n}}(\varphi_\iota).
 \ee
In the new coordinates the  Hamiltonian \eqref{stark-par} acquires the following form,
\be
\label{Hpar}
\Hstark=-\frac{2}{\xi+\eta}\left(
\gamma+
B_\xi+B_\eta
 \right)
 +\frac{\tilde {\cal I}}{\xi\eta}
+\frac{F}{2}(\xi-\eta),
\ee
where
\be
\label{Bxi}
B_\xi=\frac{1}{\xi^{\frac{N-3}{2}}}\frac{\partial}{\partial\xi}\xi^{\frac{N-1}{2}}\frac{\partial}{\partial\xi}.
\ee

We  further proceed by extending straightforwardly the steps, applied for the usual Coulomb
 system in an external field in Ref.~\onlinecite{helfrich}. Employing the following ansatz to the total wavefunction,
\be
\label{psi}
\Psi(\xi,\eta,\varphi_\iota)=\Phi_1(\xi)\Phi_2(\eta)\psi(\varphi_\iota),
\ee
we decouple the Schr\"odinger equation $\Hstark\Psi={ E}\Psi $
into three parts.  Two of them depend on $\xi$ and $\eta$, respectively,  and replace the
aforementioned equations  \eqref{R-Shr} and \eqref{Phi-Shr} for the Calogero-Coulomb system
in the spherical coordinates,
\begin{subequations}
\begin{align}
\label{sep-par1}
\left(B_\xi
      +\frac{E}{2}\xi -\frac{ F}{4}\xi^2-
      \frac{\tq(\tq+N-3)}{4\xi}
   +\lambda_1 \right)\Phi_1(\xi) & =0  ,
\\
\label{sep-par2}
    \left( B_\eta
      +\frac{E}{2}\eta +\frac{ F}{4}\eta^2-
      \frac{\tq(\tq+N-3)}{4\eta}+\lambda_2
    \right)\Phi_2(\eta)
    &= 0 .
\end{align}
Here the Coulomb charge is fractioned  into two parts,
\be
\label{gamma}
\lambda_1+\lambda_2=\gamma.
\ee
\end{subequations}
The third equation  determines the eigenstate of the algular Hamiltonian of the relative Calogero model with the eigenvalue
$\tilde I_q$ given by \eqref{Iq}.
It remains unchanged and is given by Eq.~\eqref{I-Shr}.

Evidently, in Eqs.~\eqref{sep-par1} and \eqref{sep-par2} the entire state
\eqref{psi} can be used instead of the partial wavefunctions. So, substituting  $\Phi_{1,2}\to \Psi$,
one can eliminate $E$ by subtracting  from the first equation  multiplied by $\eta$, and from the second one multiplied by $\xi$.
This manipulation together with Eq.~\eqref{gamma} yields  the eigenstate equation for the modified Runge-Lenz invariant for the
Calogero-Coulomb-Startk problem \eqref{Apar} in parabolic coordinates,
\begin{gather}
A\Psi=(\lambda_2-\lambda_1)\Psi,
\\
A=\frac{2}{\xi+\eta}\left(  \eta B_\xi - \xi B_\eta    \right)
 +\frac{\xi-\eta}{\xi\eta}\tilde{\mathcal{I}}-\frac{\xi-\eta}{\xi+\eta}\gamma-\frac{F}{2}\xi\eta.
\end{gather}
Therefore, the invariant $A$  is responsible for the  separation of the variables $\xi$ and $\eta$
in the parabolic coordinate system.
It is used instead of the angular Hamiltonian ${\cal I}$,
which separates the radial and azimuthal variables in the spherical coordinates.
The second invariant, given by the relative angular Hamiltonian ${\tilde{\cal I}}$,
is common in both cases and separates the relative angular degrees of freedom.

In the absence of the electric field, $F=0$, the fractional Coulomb charges $\lambda_i$ with $i=1,2$
take discrete values depending on the parabolic quantum numbers $n_i=0,1,2,\dots$:
\be
\lambda_i=\gamma\frac{n_i+\frac{1}{2}(\tq+1)}{n+\frac12(N-3)}.
\ee
They also specify the partial wavefunctions $\Phi_1(\xi)=\Phi_{n_1\tq}(\xi)$ and $\Phi_2(\eta)=\Phi_{n_2\tq}(\eta)$.
The principal quantum number, which characterizes the spectrum \eqref{En}, is expressed now via parabolic quantum numbers:
\be
 n=n_1+n_2+\tq+1.
 \label{nn-}
 \ee
As in the usual Coulomb problem \cite{ll}, the electric field completely removes the degeneracy in
the orbital momentum but preserves the degeneracy with respect to $q$.
Using the similarity between the Calogero-Coulomb-Stark  and $N$-dimensional Coulomb-Stark problems,
we can immediately write down the first-order energy correction caused by the weak electric field $F$
in perturbation theory:
\be
 E_n^{(1)}=
\frac32\left(n+\frac{N-3}{2}\right)(n_1-n_2)F\;.
\label{2stark}
\ee
Hence, as in the spherical case, inclusion of the Calogero term preserves the perturbative spectrum of the system, but changes its degree of degeneracy.

\section{Two-Center Calogero-Coulomb system}

\subsection{Two-center Coulomb system}
As is known, the two-center Coulomb problem is integrable in any dimension
\cite{ukr,2centerRL}.
Let us locate the two central Coulomb charges $\gamma_{1,2}$  at the points $\bm{x}_1= \bm{a}$ and $\bm{x}_2= -\bm{a}$.
The distances from them to the given space point are denoted by $r_{1}$ and $r_2$, respectively.
Then the Hamiltonian acquires  the following form:
\be
\label{two0}
\Hgamma^0=\frac{\bm{p}^2}{2} -\frac{\gamma_1}{|\bm{x}-\bm{a}|}-\frac{\gamma_2}{|\bm{ x}+\bm{ a}|}
       =\frac{\bm{p}^2}{2} -\frac{\gamma_1}{r_1}-\frac{\gamma_2}{r_2}.
\ee
The Hamiltonian \eqref{two0} possesses the same spacial symmetry as the Stark problem described by the Hamiltonian
\eqref{stark0}. It remains invariant under the $SO(N-1)$ rotations, which preserve the
line containing both charges.
Let us denote a unit vector along this line by $\bm{n}$:
\be
\bm{a}=a\bm{n},
\qquad
\bm{n}^2=1.
\ee
According to the aforementioned symmetry, the orthogonal components of the angular momentum tensor  $L^\bot_{ij}$, given by \eqref{Lbot},
are preserved. They provide the system with $N-1$ commuting integrals of motion.

The remaining $N$-th constant of motion, like in the case of the Stark system, is provided by the suitably modified
longitudinal component of the  Runge-Lenz vector \cite{2centerRL}:
\be
\label{Btwo}
A= \bm{L}^2 +  (\bm{a}\cdot\bm{p})^2 - 2 (\bm{a}\cdot\bm{x})\left(\frac{\gamma_1}{r_1}-\frac{\gamma_2}{r_2}\right).
\ee
In three dimensions, a close expression for this integral had been obtained before
 \cite{hill}. (See Ref.~\onlinecite{kryukov} for relations between both integrals.)

It is easy to see that the above quantity commutes with $L^\bot_{ij}$.

\subsection{Two-center Coulomb system with Calogero term}

Consider now this system in the presence of the inverse-square Calogero potential.
In order to construct the Hamiltonian of this system, we should replace
the momenta operators by the Dunkl momenta as in previous sections,
and then restrict the Hamiltonian to the symmetric wavefunction.
 In order to provide the system with permutation symmetry, like for the Stark Hamiltonian,
we choose the line, connecting two Coulomb charges, to be directed along the
center-of-mass coordinate \eqref{dir}. Hence, in  the Jacobi coordinates \eqref{jacobi},
the distances to the charges are given by
\be
\label{r12}
r_1=|\bm{x}-\bm{a}|=\sqrt{y^2+(x_0-a)^2},
\qquad
r_2=|\bm{x}+\bm{a}|=\sqrt{y^2+(x_0+a)^2}.
\ee
The generalized two-center Calogero-Coulomb Hamiltonian is
\be
\label{gentwo}
\Hgengamma=\frac{\bm{\pdu}^2}{2}  - \frac{\gamma_1}{r_1}-\frac{\gamma_2}{r_2}.
\ee
The particular choice of the direction of the vector $\bm{a}$ guarantees its permutation
invariance.
On the symmetric wavefunctions it gives rise to the two-center Calogero-Coulomb system
defined in the Introduction \eqref{int-two}, which we present in  a compact form,
\be
\label{two}
\Hgamma=\frac{\bm{p}^2}{2}
+\sum_{i<j}^N\frac{g(g-1)}{(x_i-x_j)^2} - \frac{\gamma_1}{r_1}-\frac{\gamma_2}{r_2}.
\ee
Like the Calogero-Coulomb-Stark  Hamiltonian, it possesses the symmetry
given by the deformed angular momentum generators perpendicular to the
predefined direction \eqref{Mbot}.

It turns  out that the  modified Runge-Lenz integral \eqref{Btwo} of the $g=0$ Hamiltonian \eqref{two0}
is extended straightforwardly to the current case:
\be
\label{Atwo}
A= \mathcal{L}_2 + a^2  p_0^2 - 2ax_0\left(\frac{\gamma_1}{r_1}-\frac{\gamma_2}{r_2}\right).
\ee
The  proof is based on the extension of the elegant trick performed in  Ref.~\onlinecite{2centerRL},
where it is shown that the Hamiltonian \eqref{two0} preserves the modified Runge-Lenz component \eqref{Btwo}.

First, we define 
the generalized Calogero-Coulomb Hamiltonians \eqref{gencoul} with the shifted central charges:
\be
\label{coul12}
\mathcal{H}^\text{gen}_{\gamma_1}=\frac{\bm{\pdu}^2}{2}    - \frac{\gamma_1}{r_1},
\qquad
\mathcal{H}^\text{gen}_{\gamma_2} =\frac{\bm{\pdu}^2}{2}    -\frac{\gamma_2}{r_2}.
\ee
Then we combine in a proper way the symmetries of both Hamiltonians, which we have already discussed,
in order to  establish the conservation of the element \eqref{Atwo} for the Hamiltonian \eqref{two}.

Each Hamiltonian \eqref{coul12} preserves the deformed angular generators, defined  with respect to their
Coulomb centers. Note that the Dunkl operators \eqref{nabla} remain unchanged under the coordinate
shift $\bm{x}\to \bm{x}\mp \bm{a}$. In turn, the deformed angular momenta
are transformed under the shift of the original point according to the standard rule. For the first Hamiltonian,
it reads as
\be
L^{\gamma_1}_{ij}=\left(x_i-\frac{a}{\sqrt{N}}\right)\pdu_j
-\left(x_j-\frac{a}{\sqrt{N}}\right)\pdu_i=L_{ij}+\frac{a}{\sqrt{N}}(\pdu_i-\pdu_j).
\ee
This provider the transformation rule for the deformed angular momentum square $\mathcal{L}_2$, defined in \eqref{M2k}:
\be
\begin{split}
\label{Mshift}
\mathcal{L}^{\gamma_1}_2=\sum_{i<j}(L^{\gamma_1}_{ij})^2
&=\mathcal{L}_2 +\frac{a}{2\sqrt{N}} \sum_{i,j }\{L_{ij},(\pdu_i-\pdu_j)\}
+\frac{a^2}{2N}\sum_{i,j}(\pdu_i-\pdu_j)^2
 \\
&=\mathcal{L}_2 -\frac{a}{\sqrt{N}} \sum_{i,j }\{L_{ij},\pdu_j\}+a^2(\bm{\pdu}^2-p_0^2).
\end{split}
\ee
The second term enters also in the expression of the longitudinal component of the
 deformed Runge-Lenz vector $\mathcal{A}_1$ \eqref{A1}.
It behaves under the coordinate shift  in the following way:
\be
\sum_{i,j}\left\{ L^{\gamma_1}_{ij},\pdu_j\right\}
=\sum_{i,j}\left\{ L_{ij},\pdu_j\right\} +\frac{a}{\sqrt{N}}\sum_{i,j}(\pdu_i-\pdu_j)\pdu_j
=\sum_{i,j}\left\{ L_{ij},\pdu_j\right\}-\sqrt{N}a(\bm{\pdu}^2 -p_0^2).
\ee
Hence, the entire component behaves under the coordinate shift as
\be
\label{Ashift}
\mathcal{A}^{\gamma_1}_1
=\frac12 \sum_{i,j}\left\{ L_{ij},\pdu_j\right\}
-\sqrt{N}a(\bm{\pdu}^2 -p_0^2)-\sqrt{N}\gamma_1\frac{x_0-a}{r_1}.
\ee
Take now a combination of these two invariants of the Hamiltonian $\mathcal{H}^\text{gen}_{\gamma_1}$,
in which the term with the anticommutator is eliminated:
\be
\mathcal{L}^{\gamma_1}_2 + \frac{2a}{\sqrt{N}} \mathcal{A}^{\gamma_1}_1
=\mathcal{L}_2 - a^2\big(\bm{\pdu}^2-p_0^2\big)-2a\gamma_1\frac{x_0-a}{r_1}.
\ee
Substituting the expressions from \eqref{coul12}, \eqref{Mshift}, and \eqref{Ashift} into the conservation condition
\be
\Big[\mathcal{H}^\text{gen}_{\gamma_1}\, ,\,\mathcal{L}^{\gamma_1}_2
+\frac{2a}{\sqrt{N}} \mathcal{A}^{\gamma_1}_1\Big]=0,
\ee
and canceling out the terms containing the $\bm{\pdu}^2$ on the right part of the commutator, we come to the equation
\be
\label{comH-AM}
\left[\frac{\bm{\pdu}^2}{2}-\frac{\gamma_1}{r_1}\, , \,
 \mathcal{L}_2+a^2p_0^2-\frac{2a\gamma_1x_0}{r_1}\right]=0.
\ee
Since the square of the Dunkl momentum commutes  with the deformed angular momentum \eqref{com-duLij},
we have:
\be
\label{com-duLsq}
[\bm{\pdu}^2,\mathcal{L}_2]=0.
\ee
This simplifies the commutator in the left side of the relation \eqref{comH-AM}, and we can
replace it by
\be
\label{comH-1}
\left[\frac{\bm{\pdu}^2}{2}\, , \, \frac{2a\gamma_1x_0}{r_1}\right]
+\left[\frac{\gamma_1}{r_1}\, , \, \mathcal{L}_2+a^2p_0^2\right]=0.
\ee
Remember now that the two Hamiltonians in Eqs.~\eqref{coul12} are
distinct from each other by the sign of the shift parameter and the values of the Coulomb coupling:
\be
\mathcal{H}^\text{gen}_{g,\gamma_2}=\mathcal{H}^\text{gen}_{g,\gamma_1}|_{a\to-a, \,\gamma_1\to\gamma_2}.
\ee
Therefore, an analogue of the equation \eqref{comH-1} for $\mathcal{H}^\text{gen}_{g,\gamma_2}$ would be
\be
\label{comH-2}
\left[\frac{\bm{\pdu}^2}{2}\, , -\frac{2a\gamma_2x_0}{r_2}\right]
+\left[\frac{\gamma_2}{r_2}\, , \, \mathcal{L}_2+a^2p_0^2\right]=0.
\ee
The sum of both equations produces  the relation
\be
\left[\frac{\bm{\pdu}^2}{2}\, , \, 2ax_0\left(\frac{\gamma_1}{r_1} -\frac{\gamma_2}{r_2}\right)\right]
+\left[\frac{\gamma_1}{r_1}+\frac{\gamma_2}{r_2}\, , \, \mathcal{L}_2+a^2p_0^2\right]=0.
\ee
Again using  the relation \eqref{com-duLsq}, we conclude that the above equation is equivalent to the
conservation condition for the modified Runge-Lenz vector's component
along the symmetry axis of the system \eqref{Atwo}:
\be
\left[ \Hgengamma\, ,\, A\right]=0.
\ee

\subsection{Separation of variables in elliptic coordinates}
Now let us show that in complete analogy with the previous case,
the two-center Calogero-Coulomb system \eqref{two} admits complete separation of variables  in  the elliptic coordinates
for $N=2,3$ and partial separation  for $N>3$.

The map from the Jacobi variables $(x_0,y)$ to the elliptic coordinates $(\xi,\eta)$ looks as follows (similar to the
usual hydrogen atom case \cite{alliluev,helfrich}):
\be
\xi=\frac{r_1+r_2}{2a},
\qquad
\eta=\frac{r_1-r_2}{2a},
\ee
where $r_i$ is the distance from the $i$-th Coulomb charge \eqref{r12}.
The relative angles $\varphi_i$ remain unchanged.
The new coordinates belong to the regions $\xi\ge 1$ and $-1\le \eta\le 1$.
The inverse transformation reads
\be
x_0=-a\xi\eta,
\qquad
y=a\sqrt{(\xi^2-1)(1-\eta^2)}.
\ee

The two-center Calogero-Coulomb Hamiltonian \eqref{two} in elliptic coordinates reads
\be
\Hgamma=\frac{1}{2a^2(\xi^2-\eta^2)}
( B_\eta-B_\xi)
 +\frac{\tilde{\cal I}(\varphi_\iota,\partial_{\varphi_\iota})} {a^2(\xi^2-1)(1-\eta^2)}
-\frac{\gamma_1}{a(\xi+\eta)}-\frac{\gamma_2}{a(\xi-\eta)},
\ee
where the operator $B_\xi$ from the kinetic energy part acquires the following form:
\be
\label{Bel}
B_\xi=\frac{1}{(\xi^2-1)^{\frac{N-3}{2}}}\frac{\partial}{\partial\xi}(\xi^2-1)^{\frac{N-1}{2}}\frac{\partial}{\partial\xi}.
\ee
Then, choosing the wavefunction,
\be
\label{Psi-ell}
\Psi(\xi,\eta,\varphi_\iota)=\Phi_1(\xi)\Phi_2(\eta)\psi(\varphi_\iota),
\ee
 we can (partially) separate
the variables in the Schr\"odinger equation $\Hgamma\Psi=E\Psi$ into the following parts:
\begin{subequations}
\begin{align}
\label{Phi1}
\left(B_\xi  -\frac{\tq(\tq+N-3)}{\xi^2-1}+2a(\gamma_1+\gamma_2)\xi+2a^2E\xi^2-\lambda\right)\Phi_1(\xi)
&=0,
\\
\label{Phi2}
\left(B_\eta  -\frac{\tq(\tq+N-3)}{\eta^2-1}+2a(\gamma_1-\gamma_2)\eta+2a^2E\eta^2-\lambda\right)\Phi_2(\eta)
&=0,
\\
\left(2\tilde{\cal I}(\varphi_\iota,\partial_{\varphi_\iota})-\tq(\tq+N-3)\right)\psi_{\tq}(\varphi_\iota)
&=0.
\end{align}
\end{subequations}

The third equation, which appears also in the spherical and parabolic cases \eqref{I-Shr},  describes the energy
eigenstates of the relative angular Calogero Hamiltonian and its spectrum,
depending on the composite quantum number $\tq$ \eqref{Iq}. In the absence of the Calogero term,
it determines the spectrum and energy states
of a  free particle system on a $(N-2)$-dimensional sphere.

Obviously, the partial states $\Phi_{1,2}$ in the first two equations depend on the energy level $E$
and the $\tq$.
The parameter $\lambda$ in the first two equations separates the variables $\xi$ and $\eta$.
It coincides with the eigenvalue of the slightly redefined Runge-Lenz invariant for the two-center Calogero-Coulomb
system \eqref{Atwo} with the Dunkl angular momentum square replaced by the doubled angular Calogero Hamiltonian,
\be
\label{Atwo'}
A= 2\mathcal{I} + a^2  p_0^2 - 2ax_0\left(\frac{\gamma_1}{r_1}-\frac{\gamma_2}{r_2}\right).
\ee
The redefinition neglects the nonessential permutational invariant $S$ term in Eq.~\eqref{L2-I}. In the $g=0$ limit,
both integrals \eqref{Atwo} and \eqref{Atwo'} become identical.
In elliptic coordinates, the above formula looks like
\be
\label{Aell}
A=\frac{1}{\xi^2-\eta^2}\left(\xi^2 B_\eta-\eta^2B_\xi\right)
+\frac{2(\xi^2+\eta^2-1)}{(\xi^2-1)(1-\eta^2)} {\tilde{\cal I}} (\varphi_\iota,\partial_{\varphi_\iota})
+2a\xi\eta\left( \frac{\gamma_1}{\xi+\eta} -\frac{\gamma_2}{\xi-\eta}    \right).
\ee

We follow the steps made above for the parabolic case.
First, use the total wavefunction $\Psi$ instead of the partial ones, $ \Phi_{1,2}$, in Eqs.~\eqref{Phi1} and \eqref{Phi2}.
Next, cancel out the energy $E$  by taking appropriate combinations of both equations.
This yields  the eigenstate equation for the modified Runge-Lenz invariant  \eqref{Aell},
\be
A\Psi=\lambda\Psi.
\ee
It separates the two equations \eqref{Phi1}
and \eqref{Phi2} from each other,
as in the absence of the Calogero potential  \cite{helfrich}.

\section{Conclusion}
In the present  paper,  we suggested new integrable generalizations
of  the rational Calogero model extended by i)
 the Coulomb potential with the Stark term and
ii)  the two-center Coulomb potential. We also found explicit
expressions of  complete sets of their constants of motion formulated
in terms of the Dunkl operators. We
demonstrated that these systems admit partial separation of variables
in parabolic and elliptic coordinates, similar to conventional Coulomb-Stark
and two-center Coulomb problems.
This allows us to conclude that the suggested  Calogero extensions preserve
the spectra of initial systems but change the degree of degeneracy.
As an illustration, we have written the first-order energy correction
of the spectrum of then Calogero-Coulomb Hamiltonian caused by an electric field.

The physical significance of the Coulomb-Stark and two-center
Coulomb systems has been well known since (at least)  the 19th century.
The importance of the Calogero model does not raise any doubts either.
The integrable synthesis, proposed in our paper,
significantly extends its area of application.
The study of some related  systems is still in order.
First of all, it is unclear  whether similar integrable systems associated with an arbitrary root  exist
and, if so, how their hidden constants of motion look.
But even in that case, we have no doubts that such generalizations
will not admit the partial separation of variables.
We intend to investigate these problems in a separate work.

\begin{acknowledgments}
A.N. thanks Alexander Turbiner for paying attention to Ref.~\onlinecite{hill}.
This work was partially supported
by the Armenian State Committee of Science Grants No.  15RF-039 and No. 15T-1C367
and by Grant No. mathph-4220 of the Armenian National Science and Education Fund based in New York (ANSEF).
The work was done   within ICTP programs  NET68,  OEA-AC-100, and  within a  project of  the Regional Training Network on Theoretical Physics
sponsored by Volkswagenstiftung under Contract No.~86260.
\end{acknowledgments}

\end{document}